\documentclass[pdflatex,sn-mathphys-num]{sn-jnl}%

\usepackage{graphicx}%
\usepackage{multirow}%
\usepackage{amsmath,amssymb,amsfonts}%
\usepackage{amsthm}%
\usepackage{mathrsfs}%
\usepackage[title]{appendix}%
\usepackage{xcolor}%
\usepackage{textcomp}%
\usepackage{manyfoot}%
\usepackage{booktabs}%
\usepackage{algorithm}%
\usepackage{algorithmicx}%
\usepackage{algpseudocode}%
\usepackage{listings}%
\usepackage{float}
\usepackage{amssymb}

\theoremstyle{thmstyleone}%

\theoremstyle{thmstyletwo}%

\theoremstyle{thmstylethree}%

\begin{document}

\title[Article Title]{LightPro: A Linear Photonic Processor with Full Programmability}

\author[1 *]{\fnm{Amin} \sur{Shafiee}}\email{amin.shafiee@colostate.edu}

\author[1]{\fnm{Zahra} \sur{Ghanaatian}}\email{zahragh@colostate.edu}

\author[2]{\fnm{Benoit} \sur{Charbonnier}}\email{Benoit.Charbonnier@cea.fr}

\author[1*]{\fnm{Mahdi} \sur{Nikdast}}\email{Mahdi.Nikdast@colostate.edu}

\affil[1]{\orgdiv{Department of Electrical and Computer Engineering}, \orgname{Colorado State University}, \orgaddress{\city{Fort Collins}, \postcode{80523}, \state{Colorado}, \country{United States}}}


\affil[2]{\orgdiv{Université Grenoble Alpes}, \orgname{CEA-Leti}, \orgaddress{\city{Grenoble}, \postcode{F38000}, \state{France}}}

\affil[*]{Corresponding Authors: Amin Shafiee and Mahdi Nikdast}

\abstract{Silicon photonics (SiPh) has enabled the integration of various optical systems across a broad range of applications, from high-speed data communication and optical I/O to energy-efficient optical computing for next-generation artificial intelligence (AI) hardware accelerators. For photonic AI acceleration, there have been numerous  implementations to improve matrix-vector multiplication (MVM), the most energy-intensive computation in deep neural network (DNN) models. However, as DNN models continue to grow in complexity and demand more scalable photonic MVM hardware, larger MVM networks with significantly more cascaded photonic devices are required. This will in turn lead to accumulated optical losses and phase and crosstalk noise, which will limit the scalability and efficiency of photonic MVM hardware. In addition, even with reduced optical losses and noise, the large size of photonic devices (e.g., Mach--Zehnder Interferometers) prohibits practical implementation of a large-scale photonic MVM hardware, as the overall footprint might be reaching or even exceeding the wafer scale.   

In this paper, we propose a novel fully programmable linear photonic processor, which we call LightPro, with improved scalability, performance, and footprint. At the heart of LightPro are compact, low-loss, and programmable SiPh directional coupler (DC) devices that deploy phase-change material (PCM) for programming the DC's splitting ratio. By thermally inducing phase transitions in the PCM, the coupling coefficient of the DC can be dynamically adjusted to achieve different splitting ratios in the device output. Building on this device foundation, we develop a neural architecture search (NAS) and pruning algorithm to optimize the architecture of the processor for performing MVM operations. Our simulation results show that LightPro achieves up to an 85\% reduction in footprint and more than 50\% improvement in power consumption. In addition, LightPro is evaluated by performing inference with weight matrices trained on MNIST and linearly separable Gaussian datasets, showing less than a 5\% drop in accuracy when scaling up the network. Prototyping results, using a commercial photonic processor (iPronics SmartLight), show LightPro's efficiency and performance (e.g., computational accuracy) compared to conventional photonic MVM hardware, demonstrating the experimental evaluation and feasibility of LightPro for next-generation photonic AI accelerators. 

}




\maketitle

\section{Introduction}\label{sec1}

As the demand for larger neural networks grows to address increasingly complex and computationally intensive tasks, artificial intelligence (AI) accelerators must deliver higher performance and accuracy while remaining energy efficient. Deep neural networks (DNNs) have become central to this effort, with applications spanning image recognition, network anomaly detection, autonomous systems, decision making, pandemic forecasting, and early cancer diagnosis, to name a few \cite{ONN_survey}. However, the continuous increase in data-heavy and compute-intensive applications calls for more complex and larger DNNs in which matrix-vector multiplication (MVM) operations are the most time- and energy-intensive operations. Yet, the most critical bottleneck today lies in the  fact that electronic DNN accelerators fail to meet the energy efficiency requirements for training and inference in emerging AI applications \cite{SiPh_codesign}. 

Silicon photonics (SiPh) has emerged as a promising solution for implementing high-speed and energy-efficient systems, offering high-performance solutions across a wide range of application domains, from ultra-fast, high-bandwidth optical interconnects in data center networks \cite{Bahadori:16} to energy-efficient optical computing in next-generation DNN hardware accelerators \cite{harris2018linear}. By performing both the communication and computation in the optical domain, integrated photonic linear processors based on SiPh can achieve up to three orders of magnitude greater energy efficiency in executing MVM operations~\cite{harris2018linear,SiPh_codesign}. There are various designs of photonic linear processors, based on both \textit{coherent} and \textit{noncoherent} multiplication. Coherent photonic multipliers, often implemented based on Mach--Zehnder interferometers (MZIs), operate at a single optical wavelength and map parameters into the signal optical phase so that optical interference performs MVM operations. In contrast, noncoherent multipliers, often implemented using microring resonators (MRRs), use multiple wavelength sources, encoding parameters into the signal transmission level to carry out multiplication \cite{tait2016microring,tait2022quantifying}. Between the two implementations, coherent multipliers eliminate the need for a multi-wavelength or comb source and offer better performance in terms of power consumption per MVM operation. Nevertheless, coherent multipliers require coherent narrow-linewidth laser sources and are susceptible to phase noise, which accumulates as the network size scales up~\cite{ghanaatian2023variation, cheng2020silicon, Clements:16,reck1994experimental,shokraneh2020diamond,De_marinis_app11136232,shen2017deep,tait2016microring}.

As the size of an MVM operation grows, more photonic devices must be cascaded within the photonic multiplier hardware to execute the computation. For example, in coherent multipliers like those based on the Clements architecture \cite{Clements:16}, each MZI incorporates two tunable phase shifters to enable reconfigurable computation by adjusting the optical phase using the phase shifters. To perform  a N$\times$N unitary transformation from input to output, $\frac{N(N-1)}{2}$ MZIs are required in the Clements network. Consequently, scaling up linear photonic processors based on existing architectures poses significant scalability challenges because of the high cumulative active power consumption, and the large footprint required by densely packed but large tunable photonic devices would result in a high phase error accumulation across the chip \cite{amin_jlt,ghanaatian2025enhanced}. In addition, with increasing chip size, maintaining signal-to-noise-ratio (SNR) becomes increasingly difficult, and the effects of process variations become more pronounced, which can degrade the SNR at the output of the network \cite{amin_jlt,ghanaatian2023variation,mirza2022characterization,ghanaatian2025bridging,ghanaatian2024mastering, harris2018linear,harris2014efficient}. Therefore, there is a critical need for a more power-efficient and compact implementation of photonic linear processors, which motivates this work.

Phase change materials (PCMs) can be co-integrated with silicon photonic devices to implement reconfigurable photonic devices and, hence, give designers additional degrees of freedom for tuning the functionality of underlying photonic devices with a relatively more compact footprint \cite{shafiee_survey, shafiee2023compact}. PCMs can reversibly switch between amorphous, crystalline, or intermediate states when triggered by an external heat source~\cite{rios2022ultra}. These phase transitions induce nonvolatile changes in both the optical and electrical properties of the material~\cite{shafiee2023survey, wuttig2017phase}, enabling dynamic, power-efficient control over device behavior. PCM-based devices support multiple optical transmission or resistance states, facilitating multi-level switching without continuous power draw. Prior efforts developed reconfigurable photonic systems---such as non-volatile phase shifters and photonic memory elements---capable of selectively modulating light propagation, optical power levels, and group delay~\cite{Shafiee2025, ocampo2024new, rios2018controlled, rios2019memory, rios2022ultra, fang2023arbitrary, shastri2021photonics,li2019fast,youngblood2019tunable,youngblood2023realization}. 

In this paper, we propose a novel coherent silicon-photonic linear processor, which we call LightPro. LightPro is designed based on programmable silicon photonic directional couplers (DCs) that integrate PCM \cite{fang2023arbitrary} (Sb$_2$Se$_3$ is considered in this paper) for full programmability. These DCs act as reconfigurable optical splitters and combiners, whose transmission characteristics can be dynamically adjusted via controlled phase transitions (i.e., changing the crystallization fraction) in the PCM~\cite{xu2019low, zhang2024tunable}. Building on this device design, we implement the LightPro architecture by cascading multiple tunable PCM-based DCs with conventional phase shifters, the orchestration of which is optimized through a proposed neural architecture search (NAS) algorithm.
Fig. \ref{Fig_overall} shows different steps for the implementation of LightPro. Starting from a dataset, we perform the Fast Fourier Transform (FFT) and use the high-frequency features for neural network training \cite{banerjee2022characterizing}. This results in complex weights between the neurons. We then take the complex weights (Fig. \ref{Fig_overall} shows a 4$\times$4 weight matrix as an example) and perform singular value decomposition to decompose the complex weights into multiplication of unitary matrices. Then, starting from the complex unitary matrices, we perform a NAS to optimize the topology of the processor as well as the reconfigurable parameters (splitting ratio of the DCs and the phase shift of microheaters), to carry out coherent MVM. LightPro enables efficient MVM in the optical domain, offering substantial 50\% and 85\% reductions in, respectively, active power consumption and chip footprint upon scaling up the MVM size. It paves the way for implementing scalable and ultra-efficient photonic AI accelerators to meet the demands of growing DNN applications. 

\begin{figure}[htbp]
\centering
\includegraphics[width=0.9\linewidth]{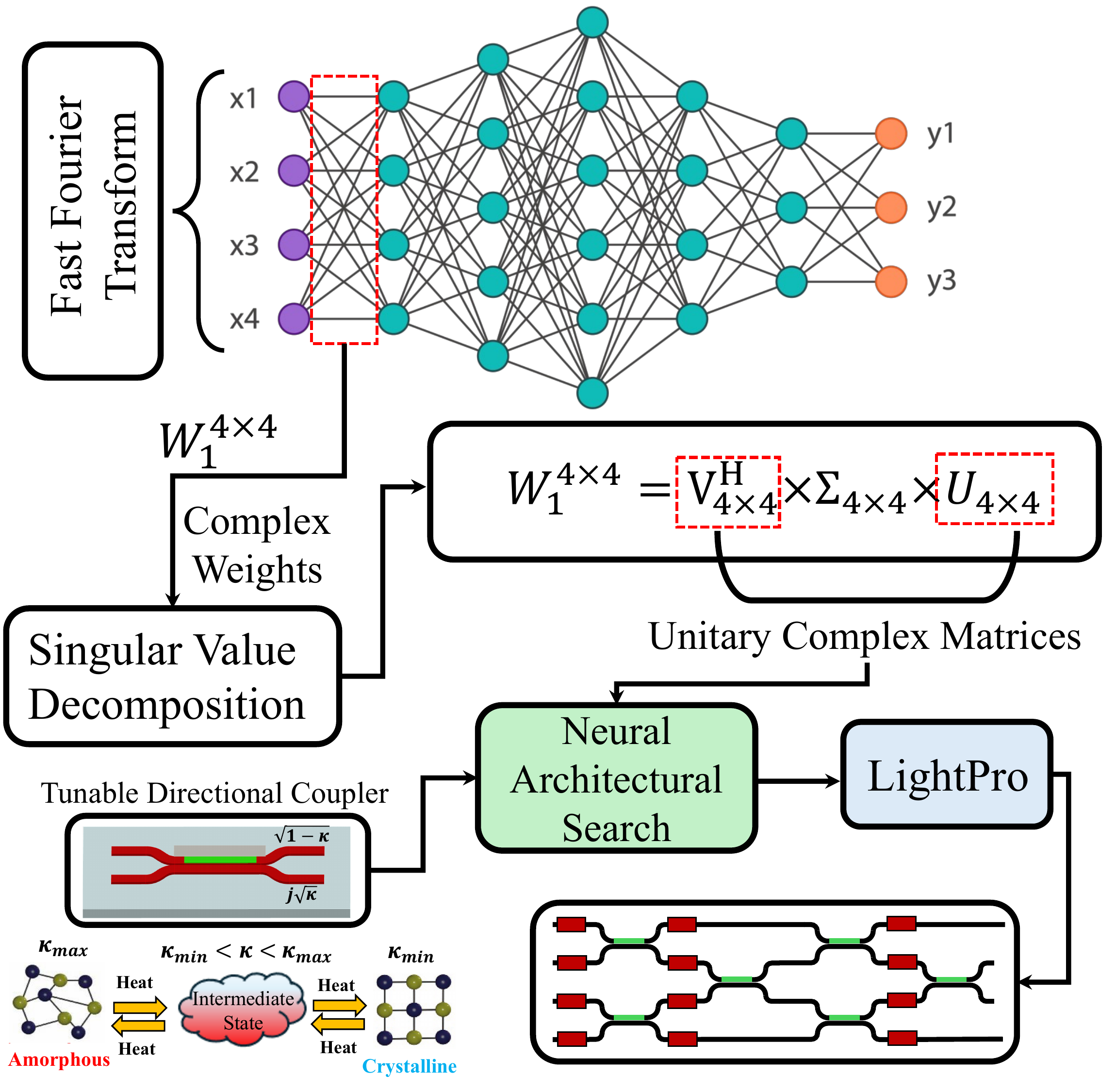}
  \caption{Working principle of LightPro. The complex weights of a DNN trained on a fast Fourier transformed dataset can be decomposed using singular value decomposition \cite{amin_jlt,banerjee2021champ}. The decomposed weight matrix can be implemented as the multiplication of two complex unitary matrices and one diagonal matrix. The complex unitary matrices can be used as a target transfer matrix in LightPro. Using the target unitary matrix, LightPro employs a NAS to optimize the network topology, leveraging columns of PCM-based tunable DCs and phase shifters to carry out MVM in the optical domain. The splitting ratio of the DCs used in the linear multiplier network can be adjusted by changing the phase state of the PCM.}\label{Fig_overall}
  \label{Fig_overall}
\end{figure}

\section{Results}\label{sec2}

\subsection{Fundamentals and Design Principle}
The coupling coefficient in silicon photonic DCs can be accurately modeled using the coupled mode theory (CMT) (see S1 and Fig. \ref{Fig_device}) \cite{cmt}. In a symmetric coupling region---where the effective refractive indices of the individual waveguides coupled in the coupling region are matched at the operating frequency---complete power transfer between the waveguides can be achieved by setting the coupling length to $L_{\pi}$ \cite{nikdast_supermod,cmt}. However, when this symmetry in the coupling region is intentionally broken (i.e. changing the width of one of the coupled waveguides, temperature change \cite{orlandi2013tunable} or using anisotropic materials \cite{yang2024fixed}), phase matching is disrupted, and the coupling efficiency will be altered. The refractive index of a PCM in both amorphous and crystalline states is significantly different from that in silicon and silicon oxide (i.e., in cladding and substrate). Therefore, when integrating a PCM on top of a waveguide in a DC, the electromagnetic phase and group delay change, and hence asymmetry can be introduced \cite{fang2023arbitrary, zhang2024tunable,xu2019low}. The schematic of the designed tunable DC that incorporates Sb$_2$Se$_3$ is shown in Fig.~\ref{Fig_device}(a). Compared to other PCM options like GST, GSST and Sb$_2$S$_3$, Sb$_2$Se$_3$ offers a good contrast in the refractive index between its amorphous and crystalline states with near-zero material loss in all phase states \cite{rios2022ultra}.
By changing the width of the waveguide and PCM such that the phase-matching condition is maintained in the amorphous state of the PCM, the coupling coefficient can be modulated by switching the PCM to its crystalline state. 

Note that silicon photonic DCs with tunable coupling coefficient can also be realized by inducing temperature change in one of the waveguides in the coupling region \cite{orlandi2013tunable} or using materials with anisotropic optical properties like Lithium Niobate \cite{yang2024fixed}. However, the active power consumption of the systems based on such a design will be high due to the essential need for maintaining active control of the DCs, especially when scaling the system. Moreover, the active tunable DCs are unable to offer a high coupling coefficient contrast compared to the one realized based on PCMs.
\begin{figure}[htp]
\centering
\includegraphics[width=1\linewidth]{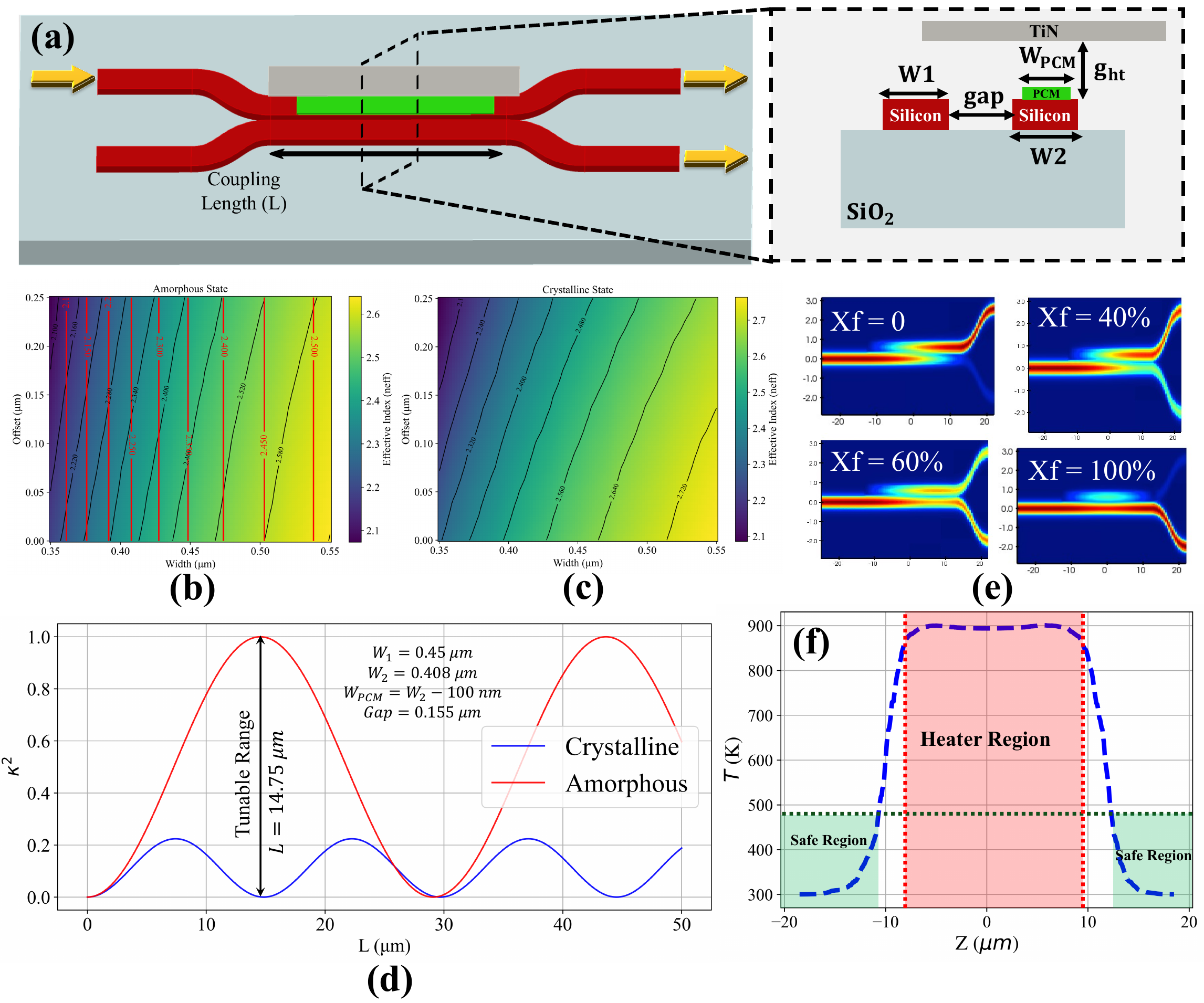}
  \caption{(a) Schematic of the tunable PCM-based DC as well as its design parameters (see the right hand side figure). A TiN microheater is used to heat the PCM and induce phase change. Exploration of effective refractive index for a Sb$_2$Se$_3$-loaded silicon waveguide with different offset and width values when the Sb$_2$Se$_3$ is in the (b) amorphous state and (c) crystalline state. (d) CMT simulation results of the coupling coefficient for an example design point where W$_{PCM}$ = offset + W$_2$, W$_1$ = 0.450, $\mu$m, W$_2$ = 0.408 $\mu$m and offset = 100~nm when the PCM is in amorphous and crystalline state. (e) The corresponding EME verification simulation for the design picked in (d) for Sb$_2$Se$_3$ has a different phase states ($X_f = 0$: Amorphous state, $X_f = 1$: crystalline state), (f) Temperature distribution of the designed heater along its length when a reset heat pulse is used to melt the Sb$_2$Se$_3$ and change its phase state to amorphous state \cite{shafiee2024programmable}.}\label{Fig_device}
  \label{Fig_device}
\end{figure}

The phase transition of the PCM is thermally induced via an integrated microheater, also illustrated in Fig.~\ref{Fig_device}(a)--right. According to an example process design kit (PDK) considered in our paper, the gap between the TiN heater and the waveguide ($g_{ht}$) is set to 600~nm with the thickness of 70~nm and minimum width of 1 $\mu$m, to ensure thermal efficiency while minimizing optical loss.  Fig.~\ref{Fig_device}(b) shows the effective refractive index of a standalone silicon waveguide with Sb$_2$Se$_3$ on top in the amorphous state at the wavelength of 1.55~$\mu$m. Note that the thickness of the PCM is kept 50~nm to ensure negligible scattering effect while maintaining high phase shift contrast between the amorphous and crystalline states \cite{shafiee2023design}. The red contours denote the effective refractive index of a standalone passive silicon waveguide without any PCM at different widths. The design points (width, offset) where the effective refractive index of the passive silicon waveguide and the silicon/Sb$_2$Se$_3$ are equal must be selected to ensure the phase matching and maximum coupling between waveguides in the coupling region in the DC.

Fig.~\ref{Fig_device}(c) shows the corresponding effective refractive index when the Sb$_2$Se$_3$  is in the crystalline state. Observe that for the range of waveguide width and offset, as the phase state of the PCM changes from the amorphous to the crystalline state or an intermediate state, the phase matching will be violated because the effective refractive index undergoes a major change. This results in a major change of the effective refractive index of supermodes in the coupling region, which leads to a change in the coupling coefficient of the DC. Next, the gap between the waveguide must be adjusted to achieve the maximum contrast in the coupling coefficient (preferably larger than 98\%) at a certain coupling length. As the aforementioned criteria are satisfied over shorter coupling length, the overall footprint of the network constructed by cascading the DCs will be reduced. Note that in this design-space exploration, W$_{PCM}$ $=$ offset $+$ W$_2$. An example design point where W$_1$= 0.45~$\mu$m, W$_2$= 0.408 $\mu$m and offset = 100~nm is selected in this work. S1 includes details of the effect of gap on the maximum coupling between the waveguides when the PCM is in the crystalline state (see Fig. S1). Note that the length of the DC will be determined by the design point we picked for the waveguide widths, offset, and gap values. The length must be chosen in a way that we achieve maximum contrast in the coupling coefficient when the phase state of the PCM changes from amorphous to crystalline state and vice versa.

Fig. \ref{Fig_device}(d) shows the coupling coefficient calculated from CMT as a function of length for the case where the gap is 155~nm. Observe that for L = 14.75~ $\mu$m, the coupling coefficient ($\kappa^2$) can change from 0--1 as the PCM phase state changes. Note that we can opt for a wider waveguide width and larger gap but this essentially comes at a cost of a larger footprint for the tunable DC. The Eigen Mode Expansion (EME) simulation for the selected design point is depicted for few phase states where the crystallization fraction changes from 0 (amorphous state) to 1 (crystalline state). Observe that as the crystallization fraction increases, the coupling of light between the waveguide in the coupling region changes from 1 to 0. The details of modeling the PCM’s optical properties in different phase states, as well as the EME simulations, are discussed in Section \ref{methods}. 

Another important design parameter affecting the density of PCM-based tunable DCs in the processor network is thermal crosstalk when microheaters are used to induce phase change in the PCM in the tunable DC structure. Ideally, the tunable DCs should be placed far enough apart to prevent thermal crosstalk from the adjacent microheaters used to change the phase state of nearby PCM cells. The 3D transient and unsteady state heat transfer simulation is performed for the integrated microheater with a length of 16 $\mu$m, a thickness of 70~nm, and a 1-$\mu$m width with $g_{ht}=600$ nm, as shown in Fig.~ \ref{Fig_device}(f). Observe that a minimum separation of 5 $\mu$m spacing from the two nearby microheaters is required to ensure that the phase state of the nearby PCMs does not change when melting the PCM ($T_m=893 ~K~(620^{{\circ}}~C)$ \cite{rios2022ultra}) on top of a DC.  Note that our prior work showed that as the $g_{ht}$ is smaller, the heat transfer to the PCM and its cooling will be more efficient \cite{shafiee2024programmable}. Moreover, as the $g_{ht}$ is smaller, the risk of the heater's melting before the PCM in the tunable DC's design melts (amorphisation) is lower. Therefore, this necessitates a more efficient heater design compared to the ones being used traditionally to implement thermo-optic phase shifters.
In the next section, we show how combining tunable DC and phase shifter columns with a neural architecture search enables optimized implementation of a complex unitary weight matrix for efficient optical-domain MVM.

\subsection{Neural Architectural Search (NAS) Algorithm}
We propose a NAS-based algorithm, the steps of which are shown in Fig.~\ref{Fig_alg}(a), to optimize the topology of LightPro's network to efficiently carry out MVM in the optical domain. Having any trained weight matrix of $W^{N\times N}$ and using singular value decomposition (SVD), we can write:

\begin{equation}
    W^{N\times M} = U^{N \times N}\Sigma^{N \times M}V^{H, M\times M}.
\end{equation}
In this formulation, $U^{N \times N}$ and $V^{H,M\times M}$ stands for $N\times N$ and $M\times M$ complex unitary matrices and can be realized through cascaded programmable DCs and phase shifters.
To realize a complex unitary weight matrix through cascaded programmable DCs and phase shifters, we introduce a progressive optimization algorithm that initiates from an empty network and incrementally constructs the architecture in an efficient and optimized manner.
 In this algorithm, Fidelity ($F$) serves as the key optimization metric, quantifying the similarity between two unitary matrices, and can be written as \eqref{fidelity}
\begin{equation}
F(U,\bar{U}) = \left( \frac{ \left| \mathrm{Tr}\left(\bar{U}^{\dagger} U\right) \right| }{N} \right)^{2},
\label{fidelity}
\end{equation}
where $\bar{U}$ denotes the deviated complex transfer matrix or the transfer matrix of the implemented network in each optimization iteration, and $U$ denotes the target complex unitary matrix calculated via SVD method using the trained weights.
Higher fidelity indicates that the matrix implemented by the optimized network is closer (i.e., more expressive) to the target unitary matrix when carrying gout MVM. As a result, the maximum Fidelity is achieved when $U=\bar{U}$ or $F=1$.

At each iteration of our proposed progressive optimization, the algorithm first searches over all the possible device types to be added to the network---either a column of traditional phase shifters based on thermo-optic effect or a column of PCM-based DCs (even or odd)--and jointly optimizes all phase shifter settings and coupling coefficients. Note that we consider device columns to support the full transformation from all inputs to all outputs of the network in each optimization iteration.

For a N$\times$N matrix, a column of phase shifters will have N phase shifters, an odd column of DCs will have $\frac{N}{2}$ DCs and an even column of DCs will have $\frac{N}{2}-1$ DCs enabling the transformation of the optical signal from inputs to outputs of the network. The device column yielding the highest Fidelity is selected. This iterative process continues until the desired fidelity threshold is achieved (i.e., 0.98--1). The models used to implement our NAS algorithm are presented in Section \ref{methods}.

The output of the progressive optimization is a general-purpose LightPro architecture capable of performing MVM on any unitary matrix. For designing linear processors where the unitary matrix is fixed (e.g., applications with stationary weight matrices or frozen trained MVM during the course of system operation), we can further improve LightPro's architecture by eliminating redundant phase shifters and DCs. Fig.~\ref{Fig_alg}(b) shows an overview of the proposed pruning technique for application-specific LightPro architecture. The pruning process starts with selecting the first phase shifter in the first column for removal, after which the remaining network parameters are re-optimized. The fidelity of the resulting pruned network is then evaluated against the fidelity from the preceding iteration (corresponding to the optimized baseline in the first iteration). A phase shifter is permanently eliminated if the fidelity degradation, $\Delta F$, is below a predefined threshold $\Delta F^t$, while the overall fidelity remains greater than 0.8. The reason for choosing the minimum fidelity threshold of 80\% in the phase shifter pruning is that our prior analysis showed that the coherent photonic linear multipliers experience less than 5\% accuracy drop when their fidelity is higher than 80\% \cite{amin_jlt,banerjee2022characterizing,mirza2022characterization,shafiee2025luxnas}. The same procedure will be performed for every phase shifter in the network to remove the redundant phase shifters while re-adjusting the remaining ones. Note that as the network scales up, a smaller $\Delta \theta_{Tol}$ must be used for removing phase shifters. As the network scales up, the number of phase shifters also increases; therefore, considering a large $\Delta \theta_{Tol}$ in phase shifter pruning, especially when scaling up the network, can lead to a significant accumulated drop in the network fidelity once the pruning is finished due to increased number of phase shifters. The accumulated drop in the pruned network's fidelity can lead to a significant drop in the pruned network's accuracy.
\begin{figure}[htp]
\centering
\includegraphics[width=1\linewidth]{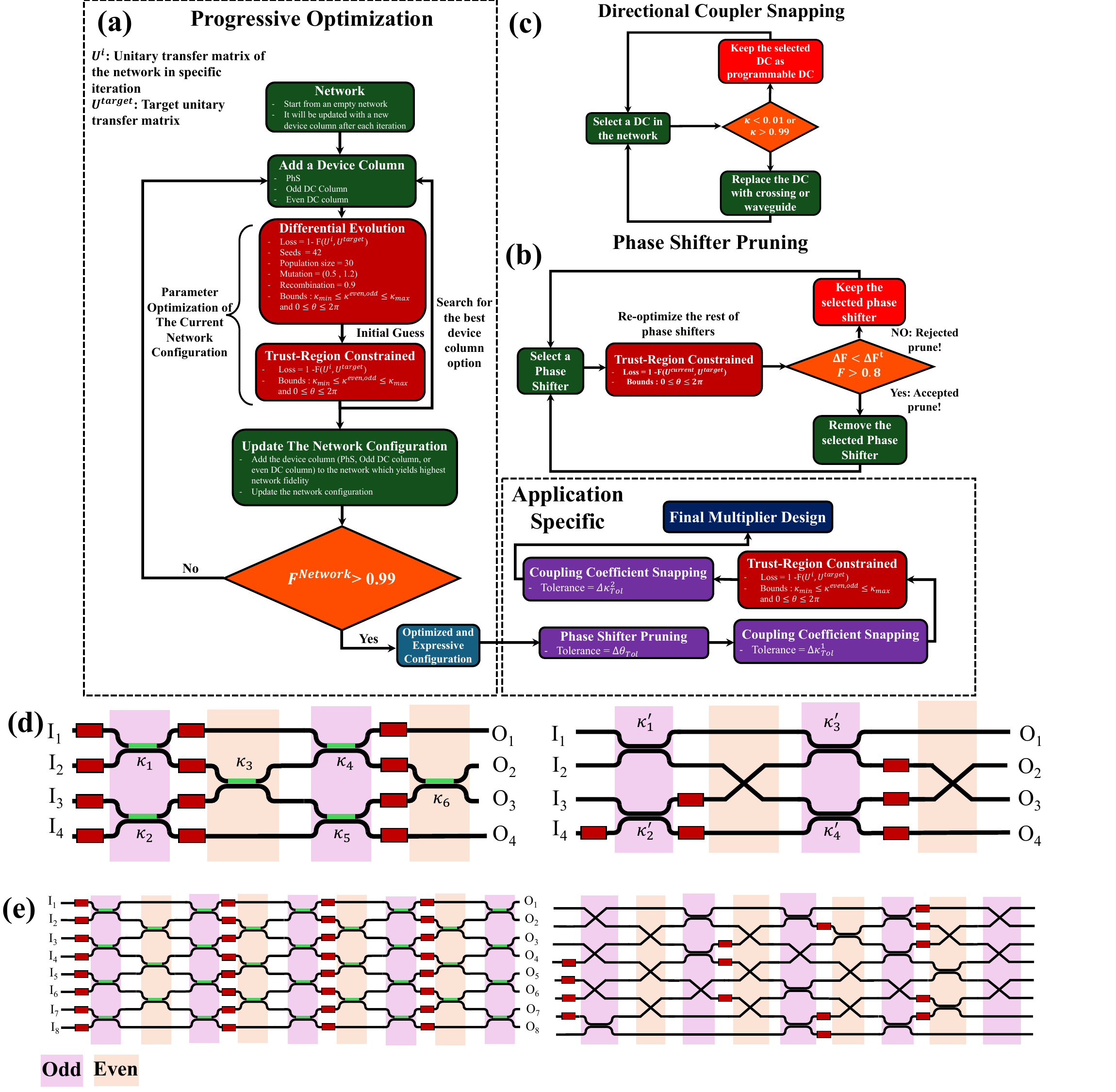}
  \caption{An overview of the proposed progressive optimization based on NAS to find the best network configuration in LightPro. Here, $\kappa_{min, max}$ can be set to be 0 and 1 for the initial progressive optimization. (b) Proposed pruning to remove phase shifters from the general-purpose LightPro network, to perform MVM for a specific application (i.e., when the unitary matrix is fixed). (c) Proposed snapping method to replace DCs with waveguide crossing or straight waveguide. (d) Left: An example of a 4$\times$4 LightPro network optimized using NAS, and right: the equivalent pruned LightPro network. (e) Left: An example of a 8$\times$8 LightPro network optimized using NAS, and right: the equivalent pruned LightPro network. In (d) and (e), the green boxes show PCM and the red ones show thermo-optic phase shifters.}\label{Fig_alg}
  \label{Fig_alg}
\end{figure}

\begin{figure}[htbp]
\centering
\includegraphics[width=1\linewidth]{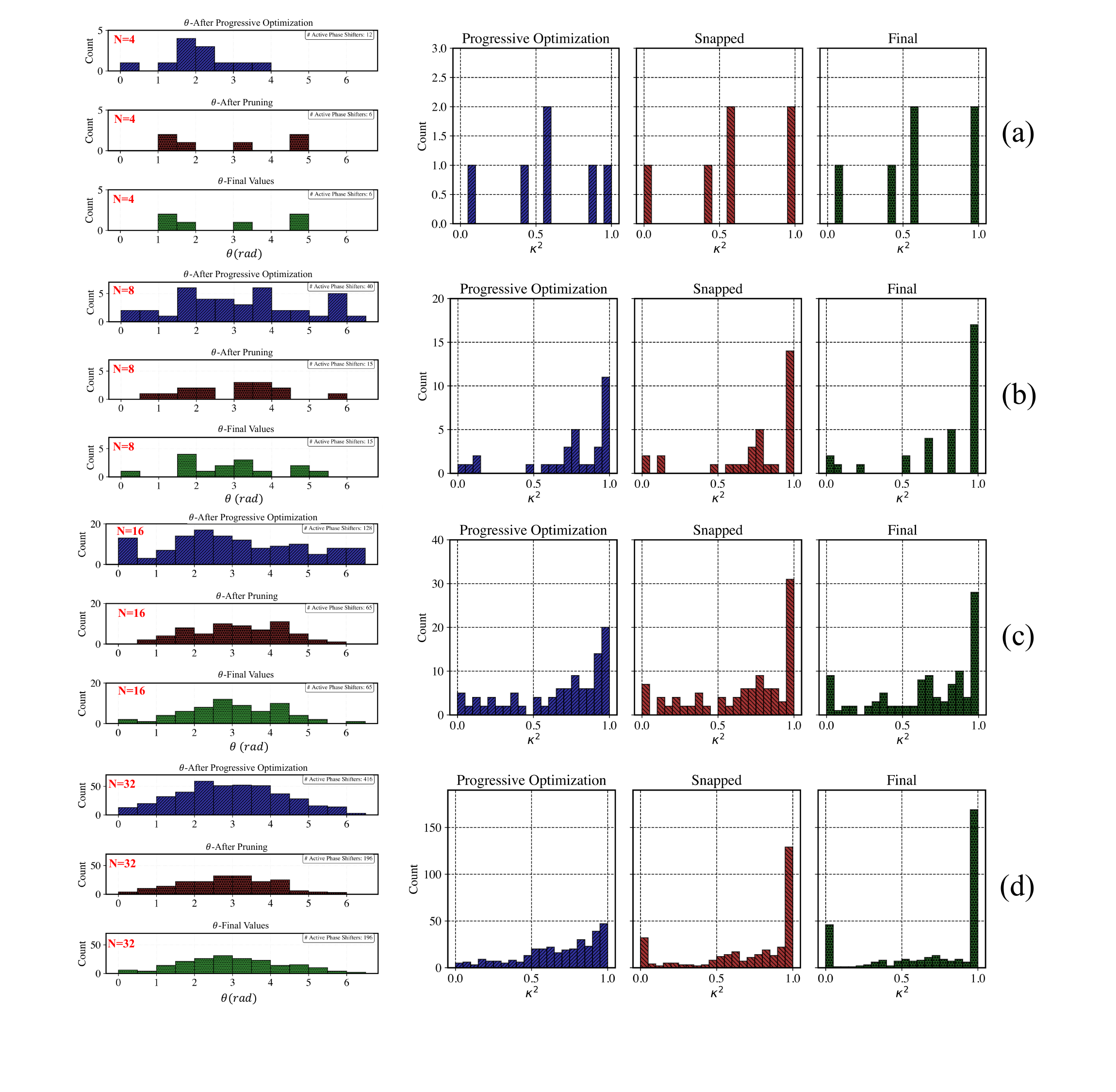}\vspace{-0.5 in}
  \caption{Phase shift and coupling coefficient distribution of the NAS optimized networks after the progressive optimization and the pruning step. The left hand side figures show the distribution of phase shifts in the network at different stages of optimization and pruning, while the right-hand side figures show the coupling coefficients after progressive optimization of the network configuration and snapping stages.}\label{Fig_sims}
  \label{Fig_sims}
\end{figure}
 After pruning phase shifters, we can snap DCs with extreme coupling coefficients (less than 0.01 and higher than 0.99 in our case to ensure a fidelity higher than 0.8 after re-adjustment of remaining parameters) and replace them with either a waveguide crossing or a straight waveguide (see Fig. \ref{Fig_alg}(c)). To do this, we perform the same procedure as the phase shifter pruning process, but without the parameter re-adjustment. The procedure begins with selecting the first DC in the first column. If the coupling coefficient $\kappa$ satisfies $1-\kappa \leq \Delta \kappa^{1}_{tol}$ or $\kappa \leq \Delta \kappa^{1}_{tol}$ and $\Delta \kappa^{1}_{tol} = 0.01$, the DC is replaced with either a waveguide crossing (for $\kappa = 1$) or a straight waveguide (for $\kappa = 0$). The reason for using $\Delta \kappa^{1}_{tol} = 0.01$ is that our analysis showed that this threshold ensures fidelity higher than 0.8 for network sizes of N$=$4, 8, 16, and 32 after final parameter re-adjustment. Following this initial snapping step, the remaining phase shifters and DCs are re-adjusted to recover the lost fidelity. Then a subsequent snapping step is performed with the smaller tolerance ($\Delta \kappa^{2}_{tol} = 10^{-3}$) to further eliminate DCs while preserving the fidelity of the network with less than 2\% fidelity drop. By replacing DCs with waveguide crossings or straight waveguides, the network exhibits improved tolerance to process-induced variations, which otherwise accumulate as crosstalk and degrade the output signal-to-noise ratio \cite{de2021photonic,amin_jlt,ghanaatian2023variation,mirza2022characterization,shafiee2025pcm}. At this stage, DCs can also be regarded as passive elements (i.e., without any PCM), since their multiplication factors are fixed and the corresponding weights remain static over time for a given application.


The motivation for opting for traditional thermo-optic phase shifters instead of PCM-based phase shifters in LightPro is the better control over the induced phase shift and the offset phase shift related to the initial phase state of the PCM-based phase shifters, which will be introduced in the network as it was shown in \cite{shafiee2023compact}. Such an offset will be challenging to characterize and calibrate in the LightPro network, especially in the pruned version when the distribution of the active phase shifters in the final network topology won't be symmetric. 

Note that although we opted for thermo-optic phase shifter for our analysis, one can still use PCM-based phase shifters in both LightPro (the general-purpose network after the progressive optimization) and Clements network. Doing so will enable zero static power consumption in both cases over the course of operation, in addition to leading to a more compact network. However, even in this case, LightPro is able to carry out the same computation with comparable programming energy related to the PCM cells in the network but with much smaller footprint area (see Fig. S3). The details on the analysis for the case where all phase shifters are also based on PCM is demonstrated in S3. Furthermore, our prior work in \cite{shafiee2025luxnas} showed that replacing the MZIs in the Clements' network with tunable DCs cannot achieve a fidelity higher than 0.75 due to the essential need for phase shifters to induce optical interference from network's input to its outputs.

\newpage

\subsection{Simulation Results}
To implement our NAS algorithm, we first trained a conventional MZI-based Clements mesh with dimensions of 4$\times$4, 8$\times$8, 16$\times$16, and 32$\times$32 on a linearly separable Gaussian dataset introduced in \cite{shokraneh2020diamond,mojaver2023addressing}. The resulting complex weights were extracted and used as the target unitary transfer matrices within our proposed NAS framework. Following this, the progressive optimization procedure was carried out (see Fig.~\ref{Fig_alg}) to determine the optimized device configurations for each network size. After convergence, a pruning step tailored to the specific trained weight matrix was applied. The resulting distributions of optimized phase shifts and coupling coefficients across different optimization stages and network sizes are shown in Fig.~\ref{Fig_sims}. Observe from Fig.~\ref{Fig_sims} left-hand-side figures for phase shift distributions, about 54\% of the phase shifters can be removed after the pruning while keeping the fidelity of the network higher than 0.8, as we will show below. As for the coupling coefficients in DCs, observe Fig.~\ref{Fig_sims} right-hand-side figures for coupling coefficient distributions that about 55\% of the DCs can be replaced by either a waveguide crossing or a straight waveguide by just re-adjusting the unpruned phase shifters and the DCs.  
An example of a 4$\times$4 and 8$\times$8 network after the progressive optimization (left-hand-side figure) and after the pruning step (right-hand-side figure) is shown in Figs.~\ref{Fig_alg} (d) and \ref{Fig_alg}(e). As can be seen, a large number of phase shifters can be removed while a large number of DCs can be replaced with either a waveguide crossing or a straight waveguide through the snapping process. 
\begin{figure}[htbp]
\centering
\includegraphics[width=0.85\linewidth]{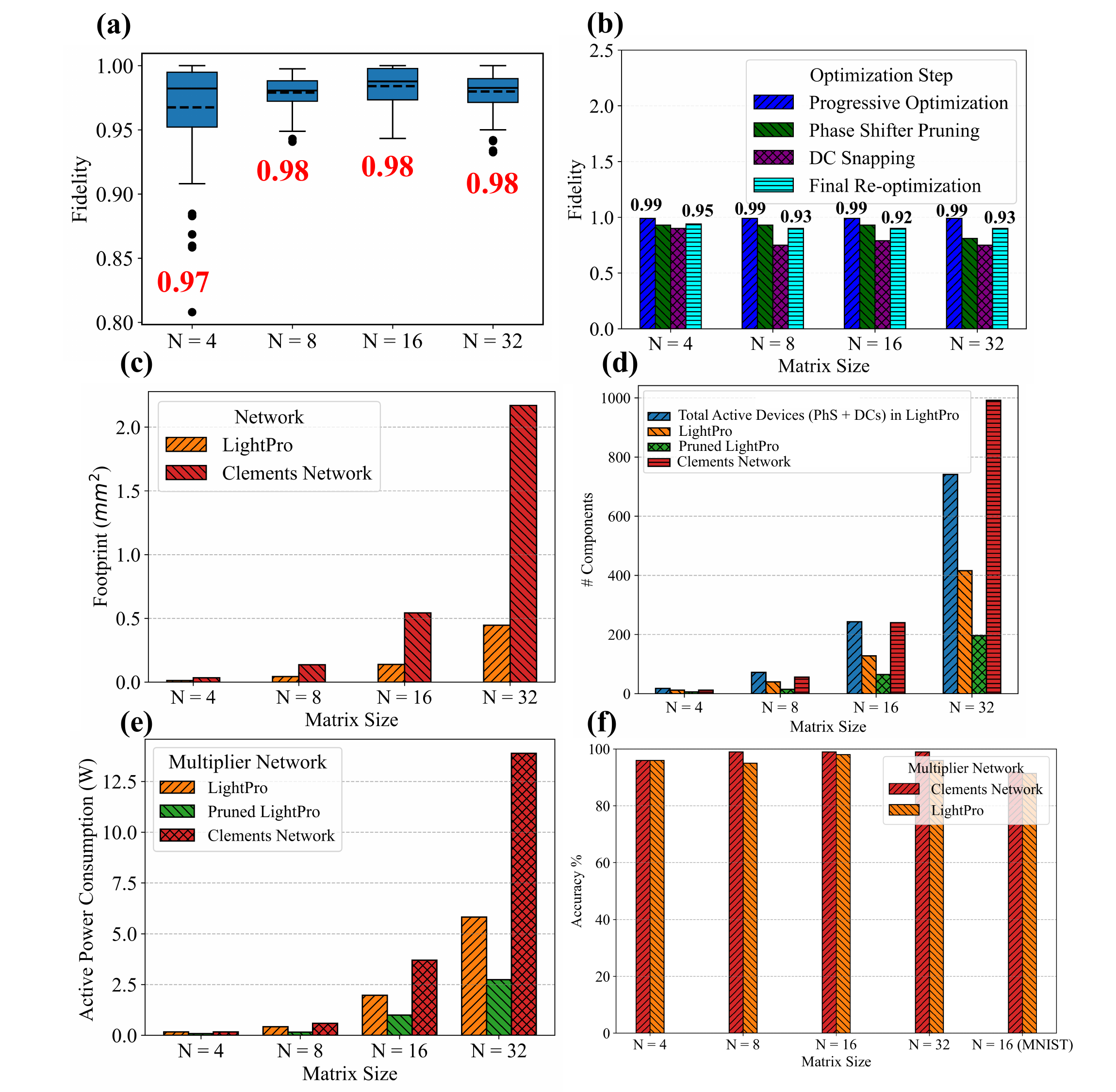}
  \caption{(a) The boxplots denoting the fidelity of the NAS optimized network of different sizes when its parameters (phase and coupling coefficients) are re-adjusted to implement 100 different random complex unitary matrix. The bold red number denotes the average achieved fidelity amongst 100 random and complex unitary matrices. (b) The fidelity of the NAS optimized network with different sizes after progressive optimization to get the optimized configuration of the network, after phase shifter pruning, after the DC snapping and final parameter re-optimization.(c) Footprint area in square millimetres for the NAS optimized network and its comparison to its equivalent MZI-based Clements network. (d) The total number of active components for the NAS optimized and pruned networks and their comparison to the MZI-based Clements network with different sizes. (e) Total active power consumption related to the phase shifters for NAS optimzied and pruned networks and their comparison with the MZI-based Clements network with different sizes when the same multiplication was implemented. Note that the phase shifters are assumed to be based on thermo-optic effect. (f) Accuracy results of pruned NAS network of different sizes when they are used to re-caclulate the accuracy of the network trained on a linearly separable Gaussian dataset. The last bar denotes the case where a 3 hidden layers with SVD configuration trained on FFT-MNIST when the original weights and the pruned weights are used to calculate the accuracy.}\label{Fig_sims_acc}
  \vspace{-0.15in}
  \label{Fig_sims_acc}
\end{figure}
To evaluate the expressivity of LightPro’s optimized architecture, i.e., its ability to represent arbitrary unitary matrices through parameter readjustment, we tested optimized configurations of different sizes (phases and coupling coefficients) against 100 random unitary matrices, without altering the network topology. For each case, we re-adjusted the parameters and recorded the fidelity between the target and implemented matrices. The results, shown in Fig.~\ref{Fig_sims_acc}(a), demonstrate that despite using a specific trained weight matrix during the NAS step to determine the optimized topology, the average fidelity across all cases remains above 0.97. This confirms the expressivity of the architecture in implementing a broad class of unitary matrices solely through phase and coupling coefficient tuning. 
Fig.~\ref{Fig_sims_acc}(b) shows the fidelity of the network trained on a target weight matrix after progressive optimization and at each pruning stage. As pruning proceeds, fidelity decreases, but final parameter re-adjustment restores up to 17\% of the lost fidelity. Fig.~\ref{Fig_sims_acc}(c) compares the footprint of the expressive configuration using tunable DCs and thermo-optic phase shifters against its equivalent MZI-based network. Note that in both network architectures, the phase shifter with the length of 200 $\mu$m and the design illustrated in S2 is considered to cover the full range of $[0,2\pi]$ phase shift (see Fig. S2).  The proposed network, combined with NAS, achieves up to an 84\% reduction in on-chip footprint while performing the same matrix–vector multiplication.  
The total number of active devices, which include phase shifters based on the thermo-optic effect and tunable PCM-based DCs in both LightPro and MZI-based Clements network is depicted in Fig.~\ref{Fig_sims_acc}(d). Observe that in both cases, the number of devices increases as the network size grows. For network sizes up to $N=$32, the total number of devices in LightPro is comparable to that of the MZI-based network, beyond which the MZI-based network requires more devices. However, given PCM-based DCs will consume no active power consumption and if we consider only the number of active phase shifters as a figure of merit, the LightPro achieves the same performance with significantly fewer phase shifters compared to the traditional MZI-based network. The total active power consumption of the phase shifters in the network after the progressive optimization, after the pruning, and for the MZI-based Clements network with different network sizes is reported in Fig.~\ref{Fig_sims_acc}(e). Observe that carrying out the same MVM operation using LightPro network reduces the active power consumption of the network, owing to the nonvolatile nature of PCMs. On top of this, further pruning the LightPro network can result in up to further $\approx$60\% active power consumption related to phase shifters in the LightPro as we scale up the network. The heat simulation results of the microheaters used as a phase shifter in both the MZI-based network and LightPro are shown in Fig. S2. 

The corresponding inference accuracy of the final pruned network, compared to the baseline MZI-based network, is shown in Fig. \ref{Fig_sims_acc}(f). Despite a fidelity drop of up to 7\% during pruning, the inference accuracy declines by less than 5\%, demonstrating the effectiveness of our pruning and re-adjustment of the network parameters method in reducing component count and active power consumption while maintaining nearly identical accuracy in matrix–vector multiplication. Fig.~\ref{Fig_sims_acc}(f) also reports the accuracy of a three-hidden-layer network trained on the more complex FFT-MNIST dataset, using both the original and the optimized/pruned weights. Even in this more demanding scenario, the pruned network incurs less than a 3\% accuracy loss relative to the baseline, further validating the robustness of the proposed method.

\begin{figure}[H]
\centering
\includegraphics[width=0.95\linewidth]{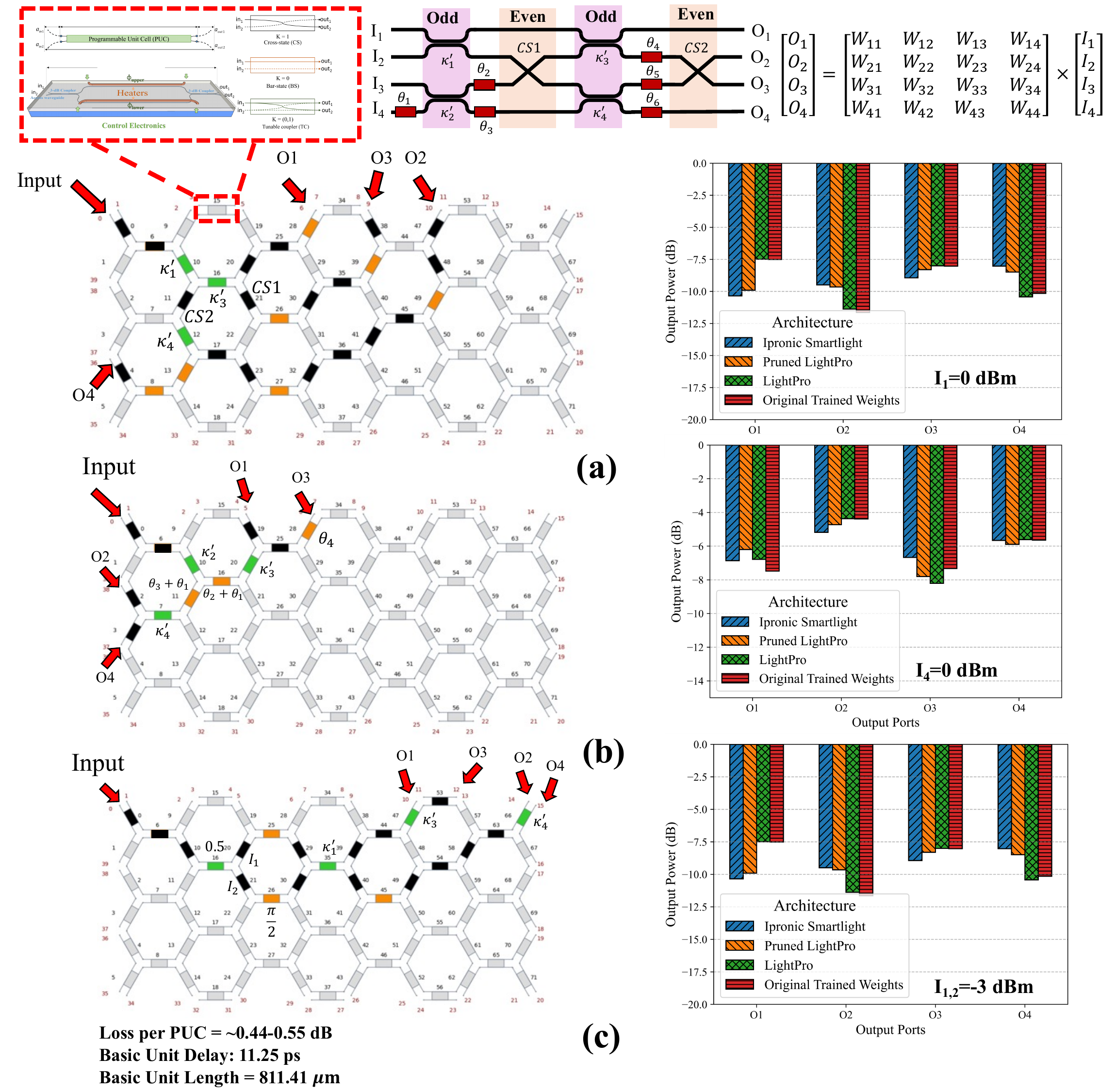}
  \caption{Experimental verification of an example 4$\times$4 LightPro network and its pruned version optimized using an example trained weight matrix. (a) Shows the verification of the case when only I$_1$ has signal, (b) Shows the verification of the case when only I$_2$ has signal, and (c) Shows the verification of the case when only I$_1$ and I$_2$ have signals. In all case studies, the output optical signal is normilized with respect to input power of the programmable mesh and the loss of the devices between input and outputs. CS1: the first waveguide crossing in the network, CS2: the second waveguide crossing in the network. }
  \label{Fig_exper}
\end{figure}

\subsection{Experimental Verification}

To experimentally verify the performance of the generated network, we prototyped a pruned $4\times 4$ network using the SmartLight iPronics processor \cite{Ipronics,bogaerts2020programmable}. The output powers from experiments were then compared with the outputs from simulations for both LightPro and when the mathematical MVM was performed using the original trained weights. The mathematical formulation of the case studies under test are illustrated in \ref{mvm}.
The programmable photonic array in iPronics SmartLight Processor consists of cascaded active MZIs arranged in a hexagonal topology, where the outputs of each MZI are connected to nearby MZIs (see Fig.~\ref{Fig_exper}). By carefully tuning the integrated phase shifters in the MZIs, a tunable DC with an arbitrary coupling coefficient can be implemented. Furthermore, the phase shifters on the MZI arms can be set to a common phase to implement a simple phase shifter, ensuring that the optical signals at both outputs of the MZI have the same phase.

Experiments were conducted using TE$_0$ polarization with $-6~\mathrm{dBm}$ optical power. Detailed specifications of the MZIs and the experimental setup are provided in Section~\ref{methods}. The first case study evaluated the network response when a signal was applied to input $I_1$ while the other inputs were off ($I_1=1$ and $I_{2,3,4}=0$). The optical power at the four outputs ($O_1$--$O_4$) was measured. Due to the limited number of coherent laser sources in the device, only one input could be analyzed at a time. The normalized output power for this scenario is shown in Fig.~\ref{Fig_exper}(a), with per-MZI losses included in the normalization for a fair comparison with the simulation results. The measured output powers (O$_1$--O$_4$) from the prototyped mesh are in good agreement with simulations for the pruned LightPro network, LightPro network, and the original mathematical MVM using the target weight matrix according to \ref{mvm}. 

\begin{equation}
\begin{bmatrix}
W_{11} & W_{12} & W_{13} & W_{14} \\
W_{21} & W_{22} & W_{23} & W_{24} \\
W_{31} & W_{32} & W_{33} & W_{34} \\
W_{41} & W_{42} & W_{43} & W_{44}
\end{bmatrix}
\begin{bmatrix}
I_{1} \\
I_{2} \\
I_{3} \\
I_{4}
\end{bmatrix}
=
\begin{bmatrix}
O_{1}=W_{11}I_{1} + W_{12}I_{2} + W_{13}I_{3} + W_{14}I_{4} = \Sigma^{N}_{i=1}W_{1i}I_{i} \\
O_{2}=W_{21}I_{1} + W_{22}I_{2} + W_{23}I_{3} + W_{24}I_{4} = \Sigma^{N}_{i=1}W_{2i}I_{i} \\
O_{3}=W_{31}I_{1} + W_{32}I_{2} + W_{33}I_{3} + W_{34}I_{4} = \Sigma^{N}_{i=1}W_{3i}I_{i} \\
O_{4}=W_{41}I_{1} + W_{42}I_{2} + W_{43}I_{3} + W_{44}I_{4} = \Sigma^{N}_{i=1}W_{4i}I_{i}
\end{bmatrix}\label{mvm}
\end{equation}

Additionally, the results were cross-validated using Ansys Lumerical Interconnect via circuit simulations of the pruned, LightPro, and MZI-based Clements networks.

The same set of analyses was performed for the case where $I_4$ is considered as the input while the other inputs are off ($I_4=1$ and $I_{1,2,3}=0$), and the output power at the four outputs was measured. As shown in Fig.~\ref{Fig_exper}(b), MZIs 16, 11, and 28 were configured as phase shifters in the programmable photonic array. The normalized outputs again match the simulation results, confirming the effectiveness of our NAS method. Finally, the input optical power in the programmable array was split to test the scenario where two inputs, $I_1$ and $I_2$ ($I_{1,2}=1$ and $I_{3,4}=0$), are active simultaneously. The two split beams were delay-matched, the configuration was applied, and the four outputs were measured. The measurement results and the test configuration are shown in Fig.~\ref{Fig_exper}(c), demonstrating excellent agreement with the simulation results.

\section{Discussion}\label{discussion}

MVM operations are at the core of DNN tasks and can be executed in the optical domain using various architectures, each with distinct trade-offs. Traditional MZI-based networks, such as the Clements mesh \cite{Clements:16}, enable complex-valued unitary transformations at a single optical frequency. However, their scalability is limited because of increased on-chip area and high power consumption from the large number of thermally or electro-optically tunable phase shifters, the number of which increases as the number of input/output increases. In this work, we propose a compact, coherent and fully programmable photonic linear processor, which we call LightPro, based on nonvolatile PCMs. Specifically, we designed a tunable DC leveraging the nonvolatile properties of PCMs that enables reconfigurable optical splitting while drastically reducing both area and power requirements. To optimize the network structure and parameters, we develop a NAS algorithm tailored for optimizing LightPro architecture by cascading columns of tunable DCs and phase shifters, and finding the best topology. Our NAS framework begins with an empty network and incrementally builds the architecture to match a target unitary transformation. Compared to a conventional Clements mesh, LightPro achieves up to an 84\% reduction in on-chip footprint while preserving single-frequency operation.

The LightPro architectures exhibit the ability to implement a large range of complex unitary matrices by just re-adjusting the networks' parameters. To evaluate the reconfigurability of the networks generated by LightPto, we performed parameter (coupling coefficients and phase shifts) re-adjustment ---without altering the network architecture---to implement 100 randomly generated complex unitary matrices. The synthesized networks achieved an average fidelity exceeding 97\% across all cases.

To further enhance power efficiency, we introduced a pruning algorithm that removes redundant phase shifters for fixed (stationary) weight matrices. The pruning method also simplifies the network by replacing unnecessary DCs with straight waveguides or crossings while readjusting the coupling coefficients of the remaining elements. This strategy reduces accumulated insertion loss and crosstalk \cite{amin_jlt}, while improving robustness to fabrication-process variations, which can otherwise degrade the extinction ratio of DCs due to their sensitivity to design deviations \cite{ghanaatian2023variation,ghanaatian2024mastering}. The pruning process achieves up to a 67\% improvement in power efficiency through removing phase shifters. In addition, we experimentally validated LightPro using a $4\times 4$ network implemented on the SmartLight Photonic Processor from iPronics, which features a programmable hexagonal MZI mesh. Multiple input combinations were tested, and the experimental results showed strong agreement with both Ansys Lumerical interconnect simulations and our in-house Python-based NAS framework.

Although Sb$_2$Se$_3$ is used as an example PCM to tune the coupling coefficient of the DCs, other PCMs exhibiting optical refractive index contrast between different phase states can also realize an asymmetric coupling region. The motivation for using PCMs to design tunable DCs is the higher contrast in the coupling coefficient they offer compared to the active tunable DCs \cite{orlandi2013tunable,yang2024fixed}. The ideal PCM for this design must exhibit extremely low material loss across all achievable phase states while providing high optical refractive contrast. Prior research indicates that phase change in Sb$_2$Se$_3$ is nontrivial, with crystallization initiating at random nucleation sites throughout the material \cite{rios2022ultra}. Therefore, detailed characterization is required to analyze the number of stable phase states between amorphous and crystalline configurations. The work in \cite{gong2024six} demonstrated that nitrogen-doped Sb$_2$Se$_3$ (N-doped Sb$_2$Se$_3$) can achieve more than 84 stable phase states, with improved optical contrast and reduced material loss. This provides approximately 0.011 resolution between different coupling coefficient states of the tunable DC. Increasing the number of stable phase states enhances the resolution for achieving various splitting ratios, giving designers greater flexibility to optimize the topology in LightPro. In this study, we assumed that every possible coupling coefficient is achievable by adjusting the material phase.

In summary, this work demonstrated a proof-of-concept of a scalable, fully programmable on-chip photonic processor for MVM using tunable photonic devices, offering significant improvements in both power consumption and footprint for accelerating complex AI tasks. The insights of this paper pave the way to implement photonic linear processors, which leverage optical interference to carry out large and complex MVM operations with significantly lower active power consumption and footprint, with more robustness to the accumulated effect of process variations.

\section{Methods}\label{methods}

\subsection{LightPro's Mathematical Models}
An $N\times N$ complex unitary matrix can be implemented by an array of cascaded device columns, including columns of phase shifters and columns of DCs, in a specific configuration ($S$) according to:
\begin{equation}\label{network}
U^{N \times N}=\prod_{(m, n) \in S} T_{i_{{m, n}}},
\end{equation} 
where $T_{i_{{m, n}}} \in \{ T_{DC}^{odd},T_{PhS}, T_{DC}^{even}\} $ and is the transformation matrix of a device column. Here, $T_{PhS}$ is the transfer matrix of a column of phase shifters connected to input ports, and $T_{DC}^{\text{odd}}$ denotes the transfer matrix of an $N \times N$ column of PCM-based DCs connected to the odd input ports ($m = 1, 3, 5, \ldots, N-1;\; n = m+1$). Similarly, $T_{DC}^{\text{even}}$ represents the transfer matrix of an $N \times N$ column of DCs connected to the even input ports ($m = 2, 4, 6, \ldots, N;\; n = m+1$). Recall that the reason for having odd and even DC columns is to have the freedom to implement any transformation from inputs to outputs of the network using tunable DCs. In this formulation,  $S$ is the network configuration and determines the order of multiplication of the transfer matrices related to different device columns, and is optimized during the progressive optimization. Moreover, an $N\times M$ complex weight matrix can be decomposed into $N\times N$ and $M\times M$ complex unitary matrices, where each can be designed using LightPro.

Considering an example 4$\times$4 LightPro network consisting of a phase shifter, odd DC and even DC columns cascaded together (i.e. see Fig. \ref{Fig_alg}(d)), $T_{PhS}$ which is a column of phase shifter with four phase shifters with phase shift values of $\theta_1$ to $\theta_4$ can be written as:
\begin{equation}
T_{PhS} = 
\begin{bmatrix}
\exp({-j\theta_1}) & 0   & 0   & 0 \\
0   & \exp({-j\theta_2}) & 0   & 0 \\
0   & 0   & \exp({-j\theta_3}) & 0 \\
0   & 0   & 0   & \exp({-j\theta_4})
\end{bmatrix}.\label{phs_mat}
\end{equation}
In addition, $T_{DC}^{odd}$ and $T_{DC}^{even}$  can be written as:
\begin{equation}
T_{DC}^{odd} = 
\begin{bmatrix}
\sqrt{1-\kappa_{1}} & j\sqrt{\kappa_{1}}   & 0   & 0 \\
j\sqrt{\kappa_{1}}   & \sqrt{1-\kappa_{1}} & 0   & 0 \\
0   & 0   & \sqrt{1-\kappa_{2}} & j\sqrt{\kappa_{2}} \\
0   & 0   & j\sqrt{\kappa_{2}}   & \sqrt{1-\kappa_{2}}
\end{bmatrix},
\end{equation}
\begin{equation}
T_{DC}^{even} = 
\begin{bmatrix}
1 & 0   & 0   & 0 \\
0   & \sqrt{1-\kappa_{3}} & j\sqrt{\kappa_{3}}   & 0 \\
0   & j\sqrt{\kappa_{3}}   & \sqrt{1-\kappa_{3}} & 0 \\
0   & 0   & 0   & 1
\end{bmatrix}.
\end{equation}
Last, the example 4$\times$4 LightPro network consisting of a column of phase shifters, a column of odd DC followed by an even DC column can be defined as:
\begin{equation}
    U^{4\times4} = T_{DC}^{even} \times T_{DC}^{odd} \times T_{PhS}.
\end{equation}
The same approach can be applied to larger networks with larger number of device columns cascaded together.

\subsection{Modeling and Simulation of Sb$_2$Se$_3$}
Experimentally extracted optical properties of Sb$_2$Se$_3$ in the amorphous and crystalline states are imported into the Ansys Lumerical Mode solver (FDE), FDTD (Finite-Difference Time Domain ), and EME (Eigen-Mode Expansion). The optical properties of the material in the intermediate states were mathematically modeled based on the Lorentz model~\cite{wang2021scheme} as:
\begin{equation}
\frac{\varepsilon_{e f f}(\lambda)-1}{\varepsilon_{e f f}(\lambda)+2}=X_{f} \times \frac{\varepsilon_{c}(\lambda)-1}{\varepsilon_{c}(\lambda)+2}+\left(1-X_{f}\right) \times \frac{\varepsilon_{a}(\lambda)-1}{\varepsilon_{a}(\lambda)+2}.
\label{main_eq}
\end{equation}
Here, $X_f$ is the crystalline fraction and takes a number between 0 and 1, illustrating the portion of the \textcolor{black}{PCM} which is in the crystalline state. Moreover, the wavelength-dependent dielectric permittivity function ($\varepsilon(\lambda)$) can be calculated as:
\begin{equation}
    \varepsilon_a = n_a^2,
\end{equation}
\begin{equation}
    \varepsilon_c = n_c^2,
\end{equation}
where $n_c$ and $n_a$ are the complex refractive indices of the \textcolor{black}{PCM}. Finally, using (\ref{main_eq}), the real and the imaginary part of the effective refractive index---which determines the phase delay and absorption of the light in a material---of a \textcolor{black}{PCM} in an intermediate (mixed) state can be estimated as:
\begin{equation}
n_{e f f}=\sqrt{\frac{\sqrt{\left(\varepsilon_{1}+\varepsilon_{2}\right)^{2}}+\varepsilon_{1}}{2}},
\label{neff}
\end{equation}
\begin{equation}
    k_{e f f}=\sqrt{\frac{\sqrt{\left(\varepsilon_{1}+\varepsilon_{2}\right)^{2}}-\varepsilon_{1}}{2}}.
    \label{nk_eff}
    \end{equation}
 In (\ref{neff}) and (\ref{nk_eff}), $\varepsilon_{2}$ and $\varepsilon_{1}$ are the real and imaginary part of $\varepsilon_{eff}(\lambda)$ in (\ref{main_eq}).

We used Ansys Lumerical Suits (FDTD, FDE, EME, HEAT, and Interconnect) for device- and circuit-level simulations and verifications. In FDTD, MODE, and EME simulations, the mesh size of 4~nm and perfectly matched layer (PML) boundary condition were used to ensure an accurate simulation of tunable PCM-based DCs.
The transient unsteady-state HEAT simulations were carried out using Lumerical HEAT solver. The time step of 1~ns and 3D simulations were used to monitor the temperature along the waveguide/PCM length when a reset pulse is used to amorphize the PCM.

To perform circuit-level simulations, we implemented different LightPro configurations in Ansys Lumerical Interconnect. The FDTD results for the DCs, and the waveguides were imported in Interconnect, ensuring accurate simulations. The optical signal delays on different paths from input to output were manually compensated, focusing on monitoring the output of the network due to optimized phase shifters and DCs. The circuit-level simulations of LightPro and its pruned versions also showed agreement with LightPro's mathematical models.

\subsection{Experimental Setup}
The programmable photonic array provided by Ipronics smartlight \cite{Ipronics} was used to verify the operation of LightPro. The internal tunable FC/APC laser source in the Ipronic device was connected to a polarization controller and then the output of the polarization controller was connected to the off-chip photodetector. Given the TE0-designed on-chip grating couplers and maximum fibre-to-chip coupling loss of 3.6 dB, we optimized the polarization to see about 5--6 dBm readout from the PD. Once the polarization of the light was calibrated, the tests for different network configurations were conducted. The MZI's extinction ratio was measured to be about 35 dB with an average MZI's insertion loss of 0.5 dB. The power consumption of the on-chip metallic heaters were 1.34--2 mW/$\pi$, which is close to our heat simulations illustrated in S2. The Ipronic smartlight programmable photonic array had in total 72 MZIs connected together with the hexagonal configuration (see Fig. \ref{Fig_exper}).

\section*{Declarations}

\subsection*{Funding Statement}
This research was supported by the National Science Foundation (NSF) under grant CNS-2046226.
\subsection*{Competing Interest}
All authors declare no financial or non-financial competing interests.

\subsection*{Authors Contribution}
A.S and Z.G carried out the multiphysic simulations and designed the optimization algorithm. A.S carried out the experimental tests and wrote the manuscript with contributions from all the authors. B.C and M.N supervised various aspects of the projects. A.S and M.N conceived the idea and oversaw the whole project. M.N provided funding for the project.
\subsection*{Data availability}
The datasets generated or analyzed during the current study are not publicly available due to conflict of interest but are available from the corresponding author on reasonable request.

\subsection*{Code availability}
The undelying code for this study is not publicly available due to conflict of interest but are available from the corresponding author on reasonable request.

\subsection*{Additional Information}
\textbf{Correspondence} and requests for materials should be addressed to Mahdi Nikdast or Amin Shafiee.









\newpage
\section*{Supplementary Information}\label{supp}

\section{Coupled Mode Theory for Asymmetric Directional Couplers} \label{CMT}

\begin{figure}[t]
\centering
\includegraphics[width=0.7\linewidth]{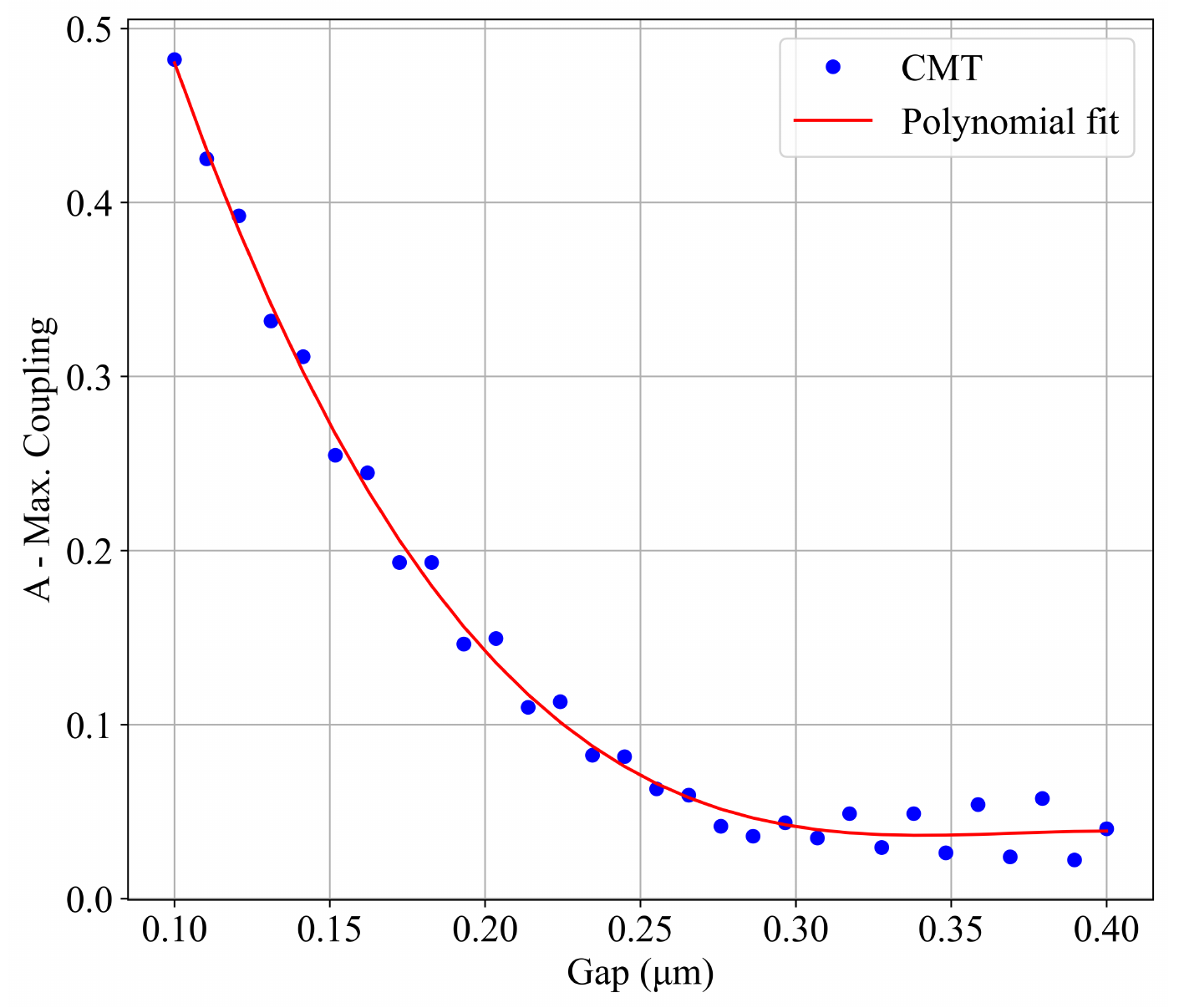}
  \caption{ Maximum coupling between the coupled waveguides in the DC when the PCM is in the crystalline state for the selected design shown in Fig.~\ref{Fig_device}}\label{Supp_m}
  \vspace{-0.15in}
  \label{Fig_exp}
\end{figure}

Coupled mode theory (CMT) was used to mathematically model and optimize the behavior of the tunable PCM-based DC. According to CMT, the asymmetric coupling region can be modeled as \cite{bayoumi2025enhanced}:
\begin{equation}
    \kappa^2=A\sin^2(\beta_c L + \phi),
\end{equation}
where $A$ denotes the maximum coupling between the two asymmetric waveguides and $\beta_c = \frac{\beta_o - \beta_e }{2}$, in which $\beta_o$ and $\beta_e$ are the propagation constants of, respectively, the odd and even supermodes. Also, $\phi$ takes into account the offset in coupling coefficient related to the input and output S-bend in the DC. As a result, the coupling coefficient of the DC when the PCM is in the amorphous state ($\kappa^2_a$) or is in the crystalline state ($\kappa^2_c$) can be modeled when the PCM on top of the waveguide is in either the amorphous or crystalline state as:
\begin{equation}
    \kappa^2_a=A\sin^2(\beta_c^a L + \phi),
\end{equation}
\begin{equation}
    \kappa^2_c=A\sin^2(\beta_c^c L + \phi).
\end{equation}
The coupling length values that result in $\Delta\kappa^2 \geq0.98$ where $\Delta\kappa = \kappa^2_a - \kappa^2_c$ can be used as the design point for our tunable PCM-based DC as changing the phase state of the PCM will toggle the $\beta_c$ between $\beta_c^a$ and $\beta_c^c$, and hence, the coupling coefficient can be tuned from 0 to 1. Moreover, the maximum coupling between the two waveguides (A) due to the asymmetric coupling region can be modeled using CMT. Thus, we can write:
\begin{equation}
    A(W_1,W_2,W_{PCM}, \lambda)\propto \Delta\beta_{1,2},
\end{equation}
where $\Delta\beta_{1,2} = \beta_1 - \beta_2^{a,c}$. Here, $\beta_1$ denotes the propagation constant of the standalone passive waveguide and $\beta_2^{a,c}$ denotes the propagation constant of the standalone PCM-loaded waveguide \cite{bent_dc}.
The maximimum coupling between the waveguide for the design point selected in this paper is shown in Fig. \ref{Supp_m}.

\section{Heat Simulation of the Thermo-Optic Phase Shifter}\label{heat}
Steady-state heat simulations were performed using Ansys Lumerical HEAT to simulate the thermo-optic phase shifters and capture the temperature distribution in a silicon-on-insulator (SOI) waveguide with a thickness of $220~\text{nm}$ and a width of $450~\text{nm}$. The heater, made of TiW alloy, was placed $2~\mu\text{m}$ above the waveguide with a length of $200~\mu\text{m}$ and a thickness of $200~\text{nm}$, following the Applied Nanotools (ANT) PDK specifications. Note that this specific design is used for both MZI-based network and LightPro network for a fair comparison, as it was fabricated before and experimentally tested according to \cite{Farhad_4by4}. The temperature profiles corresponding to each electrical power level were imported into Ansys Lumerical MODE to compute the effective refractive index variation due to the thermo-optic effect by changing the heater's power. The simulated power–phase shift relationship is presented in Fig.\ref{Supp_heat}, and these results were used to evaluate the total power consumption of the phase shifter in LightPro, pruned LightPro, and MZI-based Clements network described in Section 2.

\begin{figure}[H]
\centering
\includegraphics[width=1\linewidth]{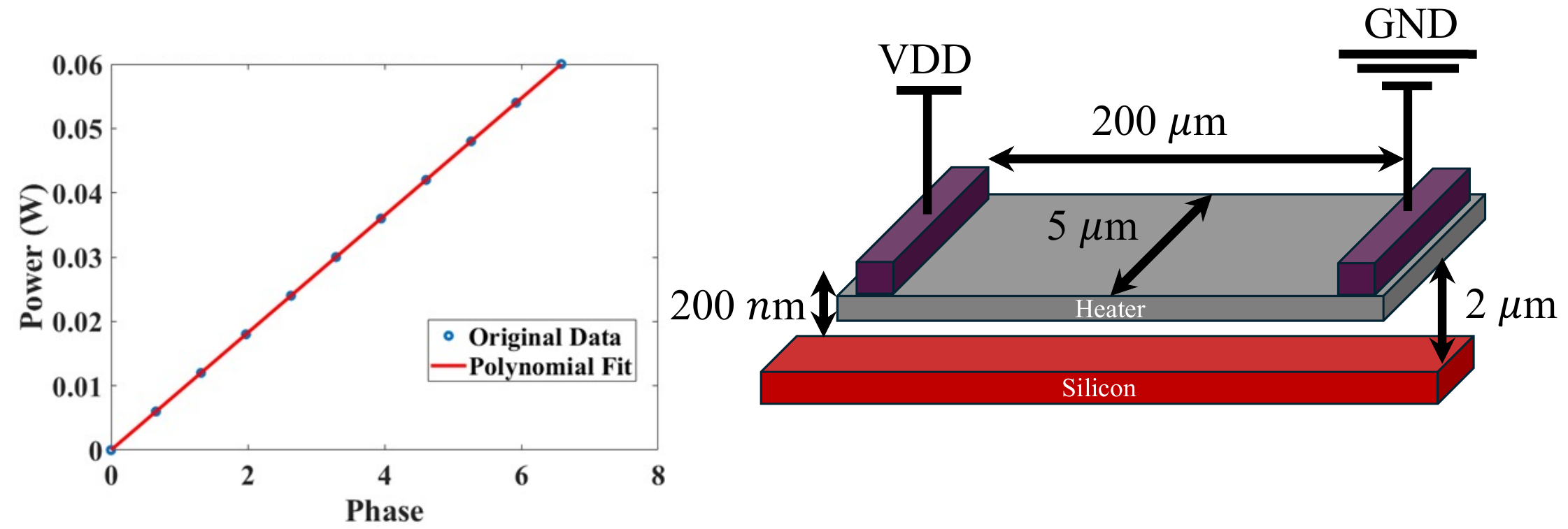}
  \caption{Power versus corresponding phase change for the 200-$\mu$m long TiN phase shifter using ANT PDK heater design.}\label{Supp_heat}
  \vspace{-0.15in}
  \label{Supp_heat}
\end{figure}

\section{All-PCM LightPro and Clements Performance Comparison}\label{all_PCM}
To compare the performance of general-purpose LightPro and Clements network of different sizes when all phase shifters are based on Sb$_2$Se$_3$ in addition to tunable DCs, we trained the Clements network of different sizes on the Gaussian dataset and then recorded the trained complex weights and the phase values related to MZIs in Clements network. The trained weights were used as a target matrix in LightPro. The design of the PCM-based phase shifter with the length of 10 $\mu$m to cover the full 2$\pi$ phase shift in the phase shifters and the MZIs with a length of 71 $\mu$m is according to our prior work in \cite{shafiee2023compact}. The heat simulation was performed to capture the case where the PCM is 50\% amorphized, 10\% amorphized and 100\% amorphized. Then an exponential curve were fitted to the data as the re-crysallization of PCM follows the Johnson-Mehl-Avrami as it was illustrated in \cite{shafiee2024programmable, wang2021scheme}. The trained phase values in Clements and optimized phases and coupling coefficient values in LightPro were then used to estimate the total programming energy of the PCMs in both networks. The results for footprint and programming energy are demonstrated in Fig. \ref{Supp_pcm}(a) and (b). Note that the design of the heater is similar to the design used for programming the tunable DCs illustrated in Section 2.
\begin{figure}[H]
\centering
\includegraphics[width=1\linewidth]{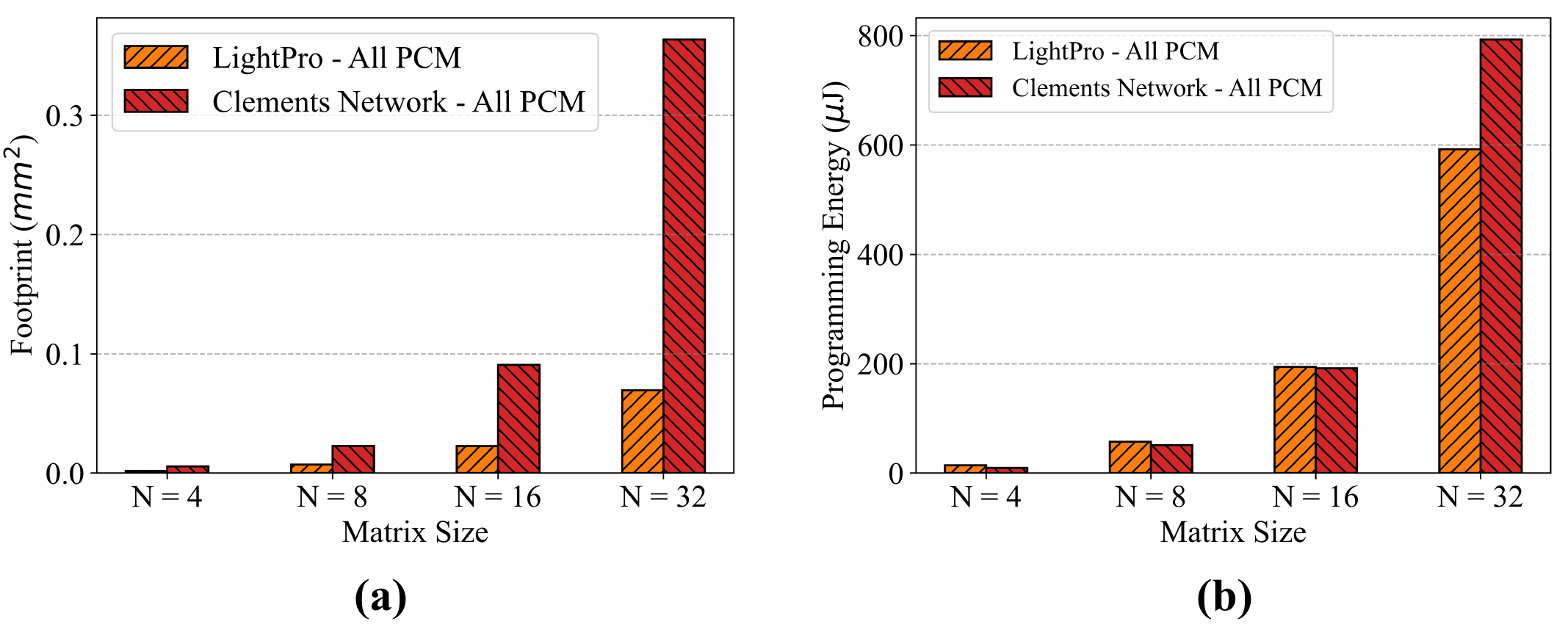}
  \caption{(a) Footprint area of the LightPro and Clements networks when all Pphase shifters are based on PCMs, (b) The programming energy for programming the PCMs in phase shifters and tunable DCs according to the trained and optimized values.}\label{Supp_pcm}
  \vspace{-0.15in}
  \label{Supp_pcm}
\end{figure}




\bibliography{sn-bibliography}

\end{document}